\documentclass[prl,twocolumn]{revtex4-1}
\usepackage{amsmath}
\usepackage{graphicx}
\usepackage{natbib}
\usepackage{color}
\begin{document}
\title{Pedagogical Materials and Suggestions to Cure  Misconceptions Connecting Special and General Relativity}
\begin{abstract}
Many professional physicists do not fully understand the implications of the Einstein equivalence principle of general relativity.  Consequently, many are unaware of the fact that special relativity is fully capable of handling accelerated reference frames.  We present results from our nationwide survey that confirm this is the case.  We discuss possible origins of this misconception, then suggest new materials for educators to use while discussing the classic twin paradox example.  Afterwards, we review typical introductions to general relativity, clarify the equivalence principle, then suggest additional material to be used when the Einstein equivalence principle is covered in an introductory course.  All of our suggestions are straightforward enough to be administered to a sophomore-level modern physics class.

\end{abstract}
\author{R. A. Pepino}
\email{rpepino@flsouthern.edu}
\author{R. W. Mabile}
\affiliation{Florida Southern College, Lakeland, FL 33801}

\maketitle

\section{Introduction}
We have found that it is fairly common for physicists who do not specialize in general relativity (GR) to believe that special relativity (SR) is incapable of modeling dynamics within accelerated reference frames.  Consequently, they may conclude that certain phenomena, such as time dilation due to acceleration, can only be described with GR.  The fact of the matter is: as long as spacetime if flat (Minkowskian), SR--not GR--is the relevant theory to use~\citep{mtw}.  In {\em Gravitation}, there is an entire chapter that discusses accelerated frames in special relativity~\citep{mtw}.  In that chapter the authors state that
``...special relativity was developed precisley to predict the physics of accelerated objects.''   Quoting {\em Spacetime and Geometry}: ``The notion of {\em acceleration}  in special relativity has a bad reputation, for no good reason"~\cite{carroll}.  We believe this misunderstanding is mainly caused by undergraduate students exposure to oversimplified discussions of the {\em twin paradox} and the {\em Einstein equivalence principle} (EEP)

To confirm the existence of this misconception within the physics community, we developed a survey, which we emailed to prestigious institutions around the United States.  This survey asked hundreds of physicists, from graduate students to professors, whether or not SR is capable modeling time dilation due to acceleration.  

In this article, we review introductions of SR and GR in six popular undergraduate modern physics textbooks~\citep{beiser,krane,thornton,harris,bernstein,taylor}.  After pointing out misstatements in three of these textbooks, we discuss their presentations of the twin paradox and the Einstein equivalence principle.  We then suggest materials that can be adopted in the classroom for these two topics that will help clarify subtle connections between SR and GR.

Specifically, with regards to the twin paradox, we encourage the usage of lines of simultaneity in worldline plots.  These lines help demonstrate to students that not not all of the time difference between frames occurs at the turnaround point.  We then introduce a straightforward example that calculates time dilation due to the acceleration of the moving twin at the turnaround point.

After students are introduced to the EEP, we suggest discussing the observations of time dilation accrued by a person in the presence of a gravitational field and another accelerating in free space by a third party observer.  The third observer would prevent students from concluding that there is a {\em global} equivalence between acceleration and gravity.  In this context, {\em geodesics}, {\em metrics}, Einstein's {\em summation convention} and the {\em Schwartzschild solution} can all be introduce into the classroom. 
Not only would the suggested materials help clarify the meaning and goals of GR, they would also introduce/reinforce concepts and tools widely used in other areas of physics.  Our suggestions are appropriate for sophomore-level physics courses and above.

Subjects like GR are esoteric, and most physicists never study the subject in great detail.  When instructors who are not familiar with the subject attempt to teach it in the classroom, they pass their own misconceptions onto the next generation.  Those students, in turn, pass their misconceptions onto the generation that follows them.  This propagation of ignorance can be easily broken at an early stage in a physics undergraduate program.

\section{Survey Questions and Responses}
The multiple choice question on our survey concerning time dilation of accelerating systems was the following:
\\
\\
{\em With regards to time dilation, what minimal level of theory is needed to solve problems involving relativistic acceleration? }
\begin{itemize}
\item Special Relativity
\item Special relativity is not capable of dealing with acceleration. General relativity is required
\item A combination of both special and general relativity are needed together
\item Cosmological relativity, because the boundary conditions of the universe are relevant
\end{itemize}
In addition to the multiple choice answer, we provided a space for free responses, which further confirmed that many were under the impression that SR was incapable of modeling dynamics within accelerated frames of reference.

Upon completion of this servey, some faculty and students disclosed their affiliated institutions.  A list of these institutions is provided below our conclusions section.

Of the several hundred surveys sent, we received 57 total responses from professionals: 38 from professors and 19 from graduate and postdoctoral students.  Excluding GR specialists, reduces the number of responses to 33 professors and 16 grad/post docs.  Although this data set is not incredibly large,
\begin{table}[h!]
\begin{tabular}{|c|c|c|c|}
\hline
{\bf Ranking} & {\bf Total Participants} & {\bf Total correct}& {\bf \% correct}\\
\hline
Faculty&33&16&48\%\\
\hline
Students&16&5&31\%\\
\hline
\end{tabular}
\caption{Survey responses from faculty and Ph.D./Postdoc students.  This data excluded physicists specializing in the field of GR.}\label{Survey Data}
\end{table}
the results are still very alarming: only 48\% of professors and 31\% of graduate and postdoctoral students were aware of the fact that time dilation due to relativitsic acceleration could be modeled solely with SR.  Of those who chose the wrong solution, 33\% of professors and 63\% of  students explicitly stated that only GR is capable of explaining time dilation for accelerating systems.

\section{Flawed Assertions in textbooks}
While discussing the twin paradox and the EEP, we found the following four inaccurate statements made in standard textbooks:
\begin{itemize}
\item ``The laws of special relativity apply only to inertial frames, those moving relative to one another at constant velocity~\citep{krane}."

\item ``...Einstein's 1905 [SR] theory applies only to non-accelerated motions... [GR] published in 1969 deals with arbitrary motions, including accelerations... when we want to deal with accelerations in [SR], we have to resort to ingenuity and approximations~\citep{bernstein}.''

\item ``Special relativity is concerned only with inertial frames of reference... [GR] goes further by including the effects of accelerations on what we observe~\citep{beiser}."

\item``...[GR] is more general than [SR] both because it includes gravity {\em and} because it focuses on noninertial, as well as, inertial reference frames*~\citep{taylor}.''
\end{itemize}
Although the asterisk in Ref.~\citep{taylor} does lead to a footnote stating ``Experts may object that even [SR] {\em can} handle noninertial frames; nevertheless, its primary focus is inertial frames''.  However, there is no ambiguity: experts {\em will} assert that accelerated frames can be treated with SR.

\section{Time Dilation in Accelerated Frames}
The key to understanding time dilation between accelerating and inertial reference frames is that for any instantaneous speed $v$ of an accelerating frame $S'$ (with respect to inertial frame $S$), there exists an inertial Lorentz transformation connecting the two frames.  This can be understood through the invariance of worldline segments.  Assuming the spacetime metric signature to be $(+,-,-,-)$:
\begin{eqnarray}
dW' & = & dW \\\nonumber
\implies c d\tau'&=&\sqrt{(c dt)^2-dx^2-dy^2-dz^2}\\\nonumber
\implies d\tau'&=&\sqrt{1-\frac{v^2}{c^2}}dt
\end{eqnarray}
for any given $v$ where $v^2=v_x^2+v_y^2+v_z^2$.  

Alternatively, this eqquation can be motivated by considering the differential form of the Lorentz transformations:
\begin{eqnarray}
dx&=&\gamma_v(dx'+v dt')\\
dt&=&\gamma_v(dt'+\frac{v}{c^2}dx') , \label{lt2}
\end{eqnarray}
where $\gamma_v=1/\sqrt{1-v^2/c^2}$ is the Lorentz factor and $c$ is the speed of light.  The differential form eliminates the need to consider changing time and spatial origins between frames.  Regardless how this discussion is motivated, for a particle stationary in $S'$, $dx'=0$.  This yields the following relationship between $t$, the time measured in $S$, to the proper time $\tau'$ of a particle at rest in $S'$
\begin{equation}
dt=\gamma_v d\tau',\label{td}
\end{equation} 
which implies
\begin{equation}
\Delta \tau'=\int_{0}^{\Delta t}\sqrt{1-\frac{v^2}{c^2}}dt .\label{propint}
\end{equation}

\section{Acceleration and the Twin Paradox}
After undergraduates learn about time dilation, they are typically exposed to the twin paradox.  Throughout this article we will refer to the stationary twin as {\em Sara} and the moving twin as {\em Mary}. This  paradox can be resolved by noting the asymmetry between the worldline trajectories of the two twins: Mary has to reverse her direction while Sara remains stationary.  In order to make the discussion as clear as possible, standard discussions involve three inertial (straight line) spacetime trajectories: one for Sara, the other two for Mary.

Some textbooks justify this inertial model by swapping Mary from an inertial frame (moving away from Sara) to another (that moves towards her)~\citep{krane,beiser,harris,thornton}.  This argument might lead some students to incorrectly conclude that, since both twins will measure the other's clock running slower than theirs before the turnaround, $100\%$ of the time dilation is accrued at the turnaround point {\em i.e.} the frame swapping itself is responsible for the time difference.

\begin{figure}[h]
\includegraphics[width=.9\columnwidth]{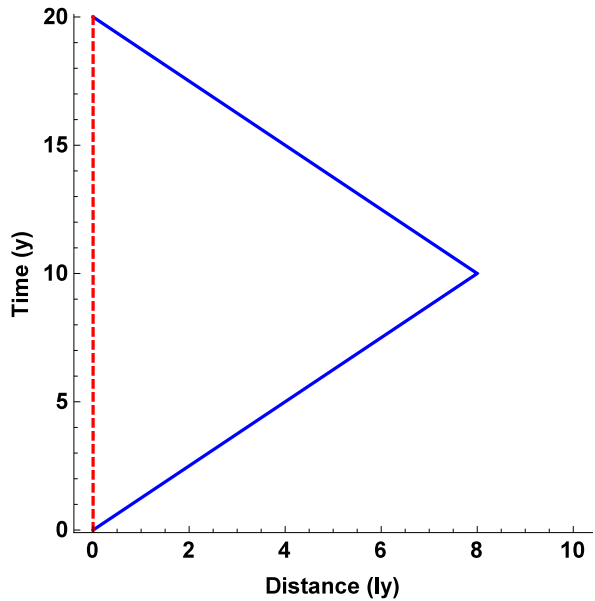}
\caption{(Color online) The standard twin paradox example.  Worldline trajectories, observed in Sara's frame, of Sara (red dashed) and Mary (blue solid) where Mary has an initial velocity of $0.8~c$, and travels to a distant planet $8~\text{ly}$ away.}
\label{std}
\end{figure}

All of the referenced books dismiss the time dilation due to acceleration at the turnaround point as negligible compared to that of the inertial dynamics~\citep{beiser,bernstein,krane,thornton,taylor,harris}.  This argument requires the magnitude of the acceleration of $S'$ be very large compared to the frame's inertial velocity measured by $S$.  However, if Mary is to survive the trip, her acceleration needs to be reasonable.  If this is the case, as we show below, the time dilation due to acceleration cannot be neglected.

\subsection{Suggested Additional Twin Paradox Material}
\subsubsection{Equal Time Plots}
\begin{figure}[h]
\includegraphics[width=.9\columnwidth]{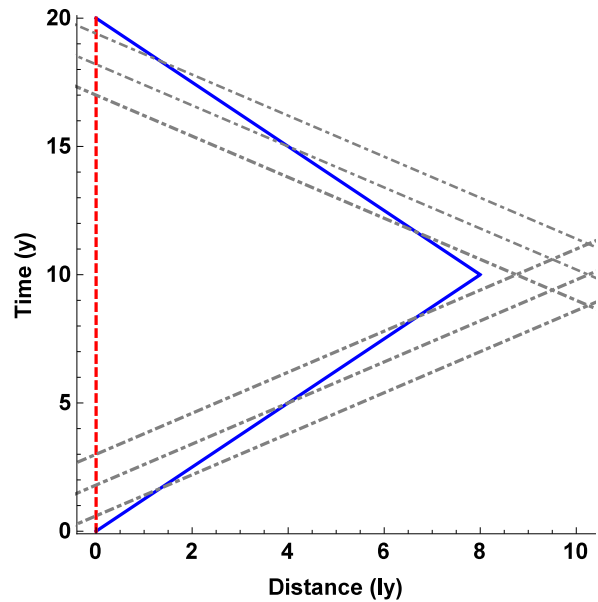}
\caption{(Color online) The standard twin paradox example.  Worldline trajectories in $S$ of Sara (red dashed) and Mary (blue solid) for Mary having an initial velocity of $0.8~c$ traveling to a distant planet $8~\text{ly}$ away.  The lines of simultaneity (gray dot-dashed) for times 1, 3, 5, 7, 9 and 11 years in $S'$ are included.
}
\label{equi}
\end{figure}

On order to help students understand that not all of the time dilation happens at the turnaround point, it is helpful to quantify and discuss the lines of simultaneity connecting Sara and Mary's clocks as observed in Sara's reference frame.  This can be done by declaring $t'$ in the standard $S\to S'$ Lorentz  transformation.  This is presented in FIG.~\ref{equi} for Mary's inertial speed is $0.8~c$ as she travels a distance of $8~\text{ly}$, which are the numbers used for the twin paradox example in Ref.~\citep{thornton}.

\subsubsection{Verification of Equation~\eqref{propint}}
The standard inertial example of the twin paradox provides the simplest mathematical computation for time dilation using Eq.~\eqref{propint}: constant $v$ implies
\begin{equation}
\Delta \tau' = \int_{0}^{t}\sqrt{1-\frac{v^2}{c^2}}dt=\sqrt{1-\frac{v^2}{c^2}}\int_{0}^{\Delta t}dt=\frac{1}{\gamma_v} \Delta t 
\end{equation}
for each leg of the trip.  This calculation also helps students see the connection between Eq.~\eqref{propint} and what they are taught from the standard Lorentz transformations.

\subsubsection{Constant acceleration}
Equation~\eqref{propint} can then be used with Mary undergoing arbitrary accelerations.  Uniform acceleration in $S$ or $S'$ are intuitive cases.

Uniform acceleration in $S'$ is a well-studied problem discussed in several textbooks~\cite{koks,carroll,rindler,ruffini}.  Although this problem could lead to discussions clarifying the true meaning of the EEP, math and analysis for this example might be difficult for early undergraduate students.  The additional times acquired for the turnaround are not negligible:  assuming ${v_i=0.8~c}$ and ${g'=9.8~\text{m/s}^2}$ leads to the additional measured time to be $2.59$ years in $S$ and $2.13$ years in $S'$.

Acceleration in $S$  is a simpler scenario, which we introduce below.  However,  constant acceleration in $S$ could lead to Mary going faster than $c$, so one must design problems such that the Lorentz factor remains real.

\begin{figure}[h]
\includegraphics[width=.9\columnwidth]{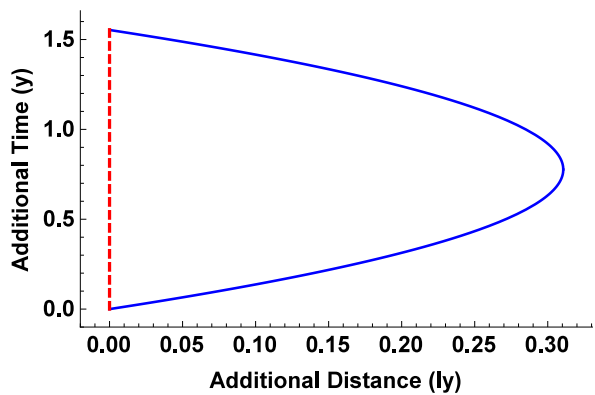}
\caption{(Color online) Kinematic trajectories in $S$.  The worldline trajectories of Sara (red dashed line) and Mary (blue solid curve) for Mary having an initial velocity of $0.8~c$ and is undergoing a constant acceleration of $a_0=9.8~\text{m/s}^2$.}
\label{accel}
\end{figure}

We consider the following problem:  at ${T_i=0}$, Sara is passed by Mary, who is traveling in the positive direction with initial speed $v_i$ while but undergoing a constant acceleration of $-a_0$ in $S$.  The worldline trajectories for this problem are depicted in Fig.~\ref{accel}.
According to Sara, Mary's trajectory is kinematic: ${v=v_i-a_0t}$.  In $S$, the added time at the turnaround is $\Delta t=2v_i/a_0$, while in $S'$, it is
\begin{equation}
\Delta \tau'=\frac{c}{g}\frac{v_i}{c}\sqrt{1-\left(\frac{v_i}{c}\right)^2}+\arcsin \left( \frac{v_i}{c}\right).\label{accT}
\end{equation}
Taking ${v_i=0.8~c}$ and ${a_0=9.8~\text{m/s}^2}$, the time measured for the turnaround is $\Delta t\simeq 1.55$ years in for Sara and ${\Delta \tau'\simeq 1.37}$ years for Mary.  These times are not negligible compared to $10$ years.  It should also be noted that, during this time interval, Mary experiences a maximum acceleration of $4.63~g$.

\section{Typical Introductions of GR}
Undergraduate physics students might only be exposed to GR in their sophomore-level modern physics course.  Coverage of GR at this level is very qualitative.  This coverage typically starts with the EEP, which can be stated as follows:
\\
\\
An observer cannot perform any {\em local} experiment to determine whether they are being uniformly accelerated or stationary in the presence of a uniform gravitational field.
\\ 
\\
Afterwards, some authors invoke the EEP to illustrate the bending of a beam of light due to gravity by drawing an equivalence to the path of light observed in an accelerating container~\cite{taylor,thornton,bernstein,beiser,krane,harris}.  However, some of these books violate the local nature of the EEP\citep{taylor,thornton,bernstein} by illustrating the light source outside of the container.

\subsection{Suggested Additional GR Material}
The term {\em local} in the EEP is very important: acceleration is not {\em globally} equivalent to gravity.  A third party observer will measure a difference between the time evolution of one person stationary in a uniform gravitational field and another undergoing uniform acceleration in free space. The rate of time dilation for the stationary indivitual is constant, while acceleration of the other implies $\gamma_v$ increases with time.  Time dilation for the stationary observer in an uniform gravitational field can be approximated using the Schwartzschild metric provided below.

\subsubsection{Suggested Introductory GR Material}
In practice, GR involves setting up Einstein's field equations for a particular scenario and then solving for the spacetime metric $g_{\mu\nu}$.  $g_{\mu\nu}$ determines how time and lengths are related to displacements~\citep{ruffini}.  Consequently, $g_{\mu\nu}$ quantifies the curvature of spacetime.  

The field equations are tensor equations which, in their compact form, are
\begin{equation}
R_{\mu\nu}-\frac{1}{2} R g_{\mu\nu}=\frac{8\pi G}{c^4}T_{\mu\nu} \label{efe}
\end{equation}
where $R_{\mu\nu}$ is the {\em Ricci tensor}, $R$ is its contraction, $G$ is Newton's gravitational constant, $c$ is the speed of light, $T_{\mu\nu}$ is the {\em energy-momentum tensor}. The subindices $\mu$ and $\nu$ each range from 0 to 3.  After defining $T_{\mu\nu}$ for a given situation, one solves for $g_{\mu\nu}$ in Eq.~\eqref{efe}.  In this form, Eq.~\eqref{efe} looks deceptively simple. Once unpacked, it yields a set of nonlinear, coupled, partial differential equations.

References \citep{taylor,thornton,bernstein,beiser,krane,harris} all mention that massive objects/energy generate curvature of spacetime.  However, none of these mainstream textbooks either present the field equations, or  mention how the curvature of spacetime is quantified.  Discussions of the field equations, and how to interpret the metrics they yield, could inspire students to learn more about general relativity.  

A geodesic is the `shortest path' to get from one point in an arbitrary space to another.  Common examples include straight lines in Cartesian space (Eq.~\eqref{cart}), or motion on the surface of a sphere.  In the context of GR, the metric relates the geodesic to the chosen coordinate system.  Below are the metric-geodesic relationships for Cartesian space ($\delta_{\mu\nu}$), Minkowski ($\eta_{\mu\nu}$) and general ($g_{\mu\nu}$) spacetime coordinates.
\begin{eqnarray}
ds^2&=&\delta_{jk}dx^jdx^k \ = ds_x^2+ds_y^2+ds_z^2\\ \label{cart}
dW^2&=&\eta_{\mu\nu} dx^\mu dx^\nu = dt^2-ds_x^2-ds_y^2-ds_z^2\\
dW^2&=& g_{\mu\nu}dx^\mu dx^\nu 
\end{eqnarray}
where $ds$ is the Cartesian length and $dW$ is, once again, the worldline segment for an arbitrary spacetime.  Here the Einstein summation convention is being used and $\{j,k\}\in\{1,2,3\}$.  

Metrics, geodesics and the Einstein summation convention show up in areas of physics ranging from classical mechanics to quantum field theory.  This makes introducing undergraduates to these topics worthwhile.  Additionally, this coverage would add a quantitative element to discussions to lessons introducing GR.

With metrics and geodesics introduced, analyses can be performed on famous solutions to the Einstein field equations such as the Schwartzchild metric: 
\begin{equation}
dW^2=\left(1-\frac{2GM}{r}\right)dt^2-\left(1-\frac{2GM}{r}\right)^{-1}dr^2-r^2d\Omega^2\label{schwartz}
\end{equation}
where $G$ is the Gravitational constant, $M$ is the mass of a spherical planet, $r$ is the distance away from the center of the planet and $\Omega$ is the solid angle.  This is the solution for a stationary spherical mass.  As stated above, this metric allows students to calculate time dilation due to gravitational fields as well as provide a quantitative introduction to black holes.

\section{Conclusions}
In this article, we have pointed out misstatements regarding SR and GR in widely-used modern physics textbooks. We have discussed the merits of covering examples of accelerated reference dynamics in SR to undergraduate physics courses.  In addition to clarifying aspects of the twin paradox, they can help eliminate common misconceptions among physicists that SR is incapable of determining time dilation due to acceleration.  We have provided a simple model of how acceleration can be taught in the classroom.  We discussed how lines of simultaneity can clarify lessons on time dilation in twin paradox example.  We suggested that, while introducing the EEP, a third party observer would help clarify its true meaning, bute also help eliminate the confusion between SR and GR.  Finally, we suggested adding basic coverage of Einstein's field equations, metrics, geodesics and the Einstein summation convention into the classroom.  Such lessons not only inspire students to learn more about GR, they add more of a quantitative component to the discussions.  The mathematical tools acquired in these lessons will also be useful for students when they take more advanced courses.

\section{Participating Institutions}\label{app}
We received surveys from several institutions.  Of the responses received, the participants who disclosed their affiliation included the following institutions:
Caltech,
Harvard University,
Iowa State University,
MIT,
Montana State University,
New Mexico State University,
North Dakota State University,
Noth Carolina State University,
Notre Dame,
Texas Tech University,
Tufts University,
University of Alabama,
University of Illinois,
University of Maryland,
University of Minnesota,
University of Nebraska,
University of North Carolina,
University of Pittsburgh,
University of South Carolina,
University of Tennessee,
University of Texas at Austin and from
Washinton University, St Louis.

\section{Acknowledgments }
We would like to thank S. Carroll and A. Hamilton for their clarifying discussions on the fundamentals of general relativity.  

\bibliographystyle{unsrt}
\bibliography{mybib}

\end{document}